\documentclass[fleqn,usenatbib]{mnras}

\usepackage{mathptmx}
\usepackage{txfonts}

\usepackage[T1]{fontenc}
\usepackage{ae,aecompl}
\usepackage{ulem} 

\usepackage{graphicx}	
\usepackage{amssymb}	

\usepackage{natbib}
\title[Pebble isolation mass in pressure maxima]{Increased isolation mass for pebble accreting planetary cores in pressure maxima of protoplanetary discs}

\author[S\'andor \& Reg\'aly]{
Zs. S\'andor$^{1,2}$ and \thanks{E-mail: Zs.Sandor@astro.elte.hu}
Zs. Reg\'aly$^2$ \thanks{E-mail: regaly@konkoly.hu}\\
$^1$Department of Astronomy, E\"otv\"os Lor\'and University, P\'azm\'any P\'eter s\'et\'any 1/A, H-1117 Budapest, Hungary\\
$^2$Konkoly Observatory, Research Centre for Astronomy and Earth Sciences, Konkoly Thege Mikl\'os \'ut 15-17., H-1121 Budapest, Hungary
}

\date{Accepted XXX. Received YYY; in original form ZZZ}

\pubyear{2020}

\begin{document}
\label{firstpage}
\pagerange{\pageref{firstpage}--\pageref{lastpage}}
\maketitle

\begin{abstract}
The growth of a pebble accreting planetary core is stopped when reaching its \textit{isolation mass} that is due to a pressure maximum emerging at the outer edge of the gap opened in gas. This pressure maximum traps the inward drifting pebbles stopping the accretion of solids onto the core. On the other hand, a large amount of pebbles ($\sim 100M_\oplus$) should flow through the orbit of the core until reaching its isolation mass. The efficiency of pebble accretion increases if the core grows in a dust trap of the protoplanetary disc. Dust traps are observed as ring-like structures by ALMA suggesting the existence of global pressure maxima in discs that can also act as planet migration traps. This work aims to reveal how large a planetary core can grow in such a pressure maximum by pebble accretion. In our hydrodynamic simulations, pebbles are treated as a pressureless fluid mutually coupled to the gas via drag force. Our results show that in a global pressure maximum the pebble isolation mass for a planetary core is significantly larger than in discs with power-law surface density profile. An increased isolation mass shortens the formation time of giant planets.
\end{abstract}
\begin{keywords}
planets and satellites: formation -- protoplanetary discs -- methods: numerical
\end{keywords}



\section{Introduction}
One of the still not entirely resolved issues of planet formation in the core accretion scenario is that the formation time of massive planets and the solid cores of giant planets may exceed the lifetime of the protoplanetary disc in the oligarchic growth models. In the original core-accretion scenario suggested by \cite{Pollack+1996}, the growth of the solid core is the result of planetesimal accretion, and the giant planet forms in about ten million years. In more elaborated physical models the long formation time is significantly reduced demonstrating the viability of the concept of planetesimal accretion \citep{Alibert+2005, Guilera+2014, Venturini+2016}. The accretion of large planetesimals is however a slow and inefficient process \citep{Fortier+2013}, therefore the demand still exists for new pathways for more rapid growth of the solid core. 

Planetary cores can grow faster if the solid particles are more tightly coupled to the gas so that nearly all of them entering the core's Hill sphere are accreted. Those particles that fulfil this criterion are called pebbles having Stokes numbers typically in the range $\mathrm{St}=0.1-1$. The growth of a solid core due to pebbles is called pebble accretion being the subject of intensive research \cite{Ormel&Klahr2010, Lambrechts&Johansen2012,Lambrechts+2014,Ida+2016,Ormel2017}, just to quote few works. 

In the usual approach to protoplanetary discs, solid particles undergo a permanent inward drift. When a solid core forms somewhere in the disc it begins to accrete pebbles and having reached a critical mass a partial gap is opened in the gas. At the outer edge of this gap a density maximum, and in connection a pressure maximum develops acting the latter as a dust trap cutting the influx of pebbles to the growing core. In this case, the growing planetary core reaches its pebble isolation mass that can be around tens of Earth masses depending on the disc model and physical phenomena considered \citep{Lambrechts+2014, Bitsch+2018, Ataiee+2018}. 
It is noteworthy that, even pebble accretion is an inefficient process because to form a $10 M_\oplus$ core starting from a Moon-sized embryo $100 M_\oplus$ mass of pebbles should flow through the orbit of the core due to their steady inward drift \citep{Morbidelli+2015, Bitsch+2019}. 

The efficiency of pebble accretion is highly improved, however, if planet formation takes place where a dust and a planet migration trap are close to each other. The increasing number of observed ring-like structures in submillimeter dust emission on high-resolution ALMA images of protoplanetary discs support the existence of such places \citep{ALMAPartnership2015, Andrews+2010, Dullemond+2018, Long+2018}. These dusty rings could be formed in global density and pressure maxima of protoplanetary discs. Already before these observations, it has been assumed in a series of theoretical works, that global pressure maxima in protoplanetary discs might be preferential places for planet formation \citep{Lyra+2008, Lyra+2009, Sandor+2011, Regaly+2012, Regaly+2013, Guilera&Sandor2017, Guilera+2020, Morbidelli2020}. A pressure maximum collects dust particles that grow further to pebble sizes triggering the streaming instability leading to the formation of larger planetesimals or even small planetary cores. A pressure maximum usually develops in connection with a density maximum in the gas which acts as a planet trap stopping the type I migration of the growing cores. When the density and the corresponding pressure maximum are generated at the water snowline, for instance, the dust trap is very close to the planet trap thus the core formation either by pebble accretion \citep{Guilera&Sandor2017} or hybrid (pebble + planetesimal) accretion \citep{Guilera+2020} is very efficient, and a giant planet forms well within the lifetime of the disc.  

Based on the above considerations it can be assumed that global pressure maxima can play important role in planet formation as they efficiently trap solid particles and shorten the formation time of planets. In this work, we study pebble accretion in a generic global pressure maximum of a protoplanetary disc. Our aim is to investigate what is the pebble isolation mass of a solid core that forms and grows in a global pressure maximum by pebble accretion. 

\section{Physical model}

In this section, we describe our disc model in which a steady-state surface density maximum is generated in gas. We call this surface density maximum {\it generic} because we do not specify in a very detailed way the physical phenomenon behind its formation. Due to the equation of state of the disc's gas, $P=\Sigma_\mathrm{gas} c_s^2$, we assume that a surface density maximum generates a pressure maximum. In what follows, when mentioning a pressure maximum, we always consider a surface density maximum associated with it.

The formation of a pressure maximum in a protoplanetary disc is usually thought to be the result of a change in the disc's effective viscosity. To avoid the disc's viscous evolution, and to focus only on the combined effect of gap opening and accretion of solids by a planetary core, we consider a steady-state disc. Similarly to the prescription of \cite{Ataiee+2014}, the background surface density of the gas $\Sigma_\mathrm{g}^\mathrm{bg} $is superimposed with a Gaussian:
\begin{equation}
\Sigma_\mathrm{g}=\Sigma_\mathrm{g}^\mathrm{bg} \left(1+A e^{-\frac{(R-R_0)^2}{\Delta R^2}}\right).
\label{eq:initgas}
\end{equation}
Considering the kinematic viscosity in the form
\begin{equation}
    \nu = \nu_\mathrm{bg} \left(1+A e^{-\frac{(R-R_0)^2}{\Delta R^2}}\right)^{-1},
\end{equation}
one can see that the condition for a steady-state gas disc is obviously fulfilled as
\begin{equation}
\nu\Sigma_\mathrm{g}=\nu_\mathrm{bg}\Sigma^\mathrm{bg}_\mathrm{g} = \mathrm{const}.
\end{equation}
In the above formulae, the maximum of the Gaussian is at $R_0$, and the quantities $A=0.5$ and $\Delta R=0.1$ adjust the amplitude and width of Gaussian, respectively. For the background surface density of gas we use a constant profile with $\Sigma_\mathrm{g}^\mathrm{bg} = 10^{-3}$ and the background viscosity is parametrized as $\nu_\mathrm{bg}=\alpha c_\mathrm{s} H$, where $\alpha=10^{-3}$. To obtain a constant $\nu_\mathrm{bg}$ both $c_\mathrm{s}$ and $H$ are evaluated at $R=1$.

Our numerical simulations are performed in a 2D locally isothermal flat disc with a constant aspect ratio $h=0.05$ in which case the radial computational domain extends between 0.5 and 2 dimensionless units divided logarithmically by 512 grid points, while the azimuthal grid is divided to 1024 sectors. Setting the maximum of $\Sigma_\mathrm{g}$ to $R_0=1.01503$, the pressure maximum is located at $R_\mathrm{pmax}=1.0$, where we initially place the growing core revolving in a circular orbit. The location of the pressure maximum $R_\mathrm{pmax}$ is calculated by equating to zero the derivative of the equation of state $P(R)=\Sigma_\mathrm{gas}(R) c_s^2(R)$ with respect to $R$, and solving this equation for $R$. The core's dimensionless mass is initially $q_0=m_0/M_*=3\times 10^{-6}$, being $m_0$ the core's, and $M_*$ the star's physical masses. We note that the pressure maximum may not coincide with the zero torque position, that according to the torque analysis by \citet{Morbidelli2020} is located in a wider domain around the gas surface density maximum depending on the disc's thermal property. To check how the pebble isolation mass depends on the position of the core, we additionally place the core to the surface density maximum being at a larger distance from the star than the pressure maximum, and symmetrically to this position, to a smaller distance from the star than the pressure maximum. To keep simulations simple, we do not allow the growing core to migrate since its migration is also influenced by the recently identified effects such as the heating torque \citep{BenitezLlambay+2015, Masset2017}, or the torque from pebbles \citep{BenitezLlambay&Pesah2018,Regaly2020}. Due to the high complexity of the planet--gas--pebble interactions, we will address the detailed investigation of this problem in a future work. In our simulations initially a uniform dust surface density $\Sigma_\mathrm{d}^0=\epsilon\Sigma^0_\mathrm{gas}$ is used, where $\epsilon = 10^{-2}$ is the disc metallicity. Solid particles are treated as a pressureless fluid that is coupled to the gas component of the disc through the aerodynamic drag. This coupling is mutual, as the backreaction of dust to gas is also taken into account. Dust particles are assumed to be grown up to pebble sizes. We consider two populations of solids consisting of pebbles characterised with only one Stokes number that is fixed either to $\mathrm{St}=0.1$ or to $\mathrm{St}=1$. Pebbles do not evolve in time during the simulations. The equations of gas and dust evolution, and the initial and boundary conditions are described in \citep{Regaly2020}. Simulations are performed by the modified GFARGO code \citep{Regaly2020}, in which a new dust module has been implemented.

The pressure maximum generated near the surface density maximum efficiently collects solid material toward its radial position $R_\mathrm{pmax}$, where the planetary core is orbiting in a pebble rich environment. At the onset of our simulations we change the mass of the planetary core gradually with a tapering time of 10 orbital periods until reaching $m_0=3\times 10^{-6}$ (that is $1M_\oplus$ when $M_*=1M_\odot$). Having reached the above initial mass, the core begins to increase its mass solely by accreting pebbles. Since we treat pebbles as fluid, we apply the same prescription for the accretion of pebbles as for the gas accretion given by \cite{Kley1999}, while for simplicity, we neglect gas accretion. The method is discussed in detail in \citep{Regaly2020}. The efficiency of pebble accretion is characterised with $0\leq \eta\leq 1$, where 0 is for no accretion, and 1 for full accretion. The mass tapering is applied for the smooth accommodation of $\Sigma_\mathrm{g}$ to the growing core, on the other hand, a significant amount of pebbles is already accumulated in the pressure maximum forming an initial mass reservoir until the planetary core begins to accrete the pebbles accumulated. This setup is reasonable, because the streaming instability that leads to the formation of large planetesimals, also requires a certain amount of solids to be triggered. We note that the streaming instability transforms pebbles to planetesimals at a certain rate, however, in this work we do not consider this effect that slightly decreases the efficiency of pebble accretion, because we only intend to study the maximum mass that a core could achieve through pebble accretion. The tapering time in our simulations has been set arbitrary to 10 orbital periods of the core leading to the formation of an initial mass reservoir around the pressure maximum. To explore how this initial mass reservoir affects the pebble isolation mass, we have also applied two additional tapering times being 5 and 20 orbital periods of the core.

\begin{figure}
    \includegraphics[width=\columnwidth]{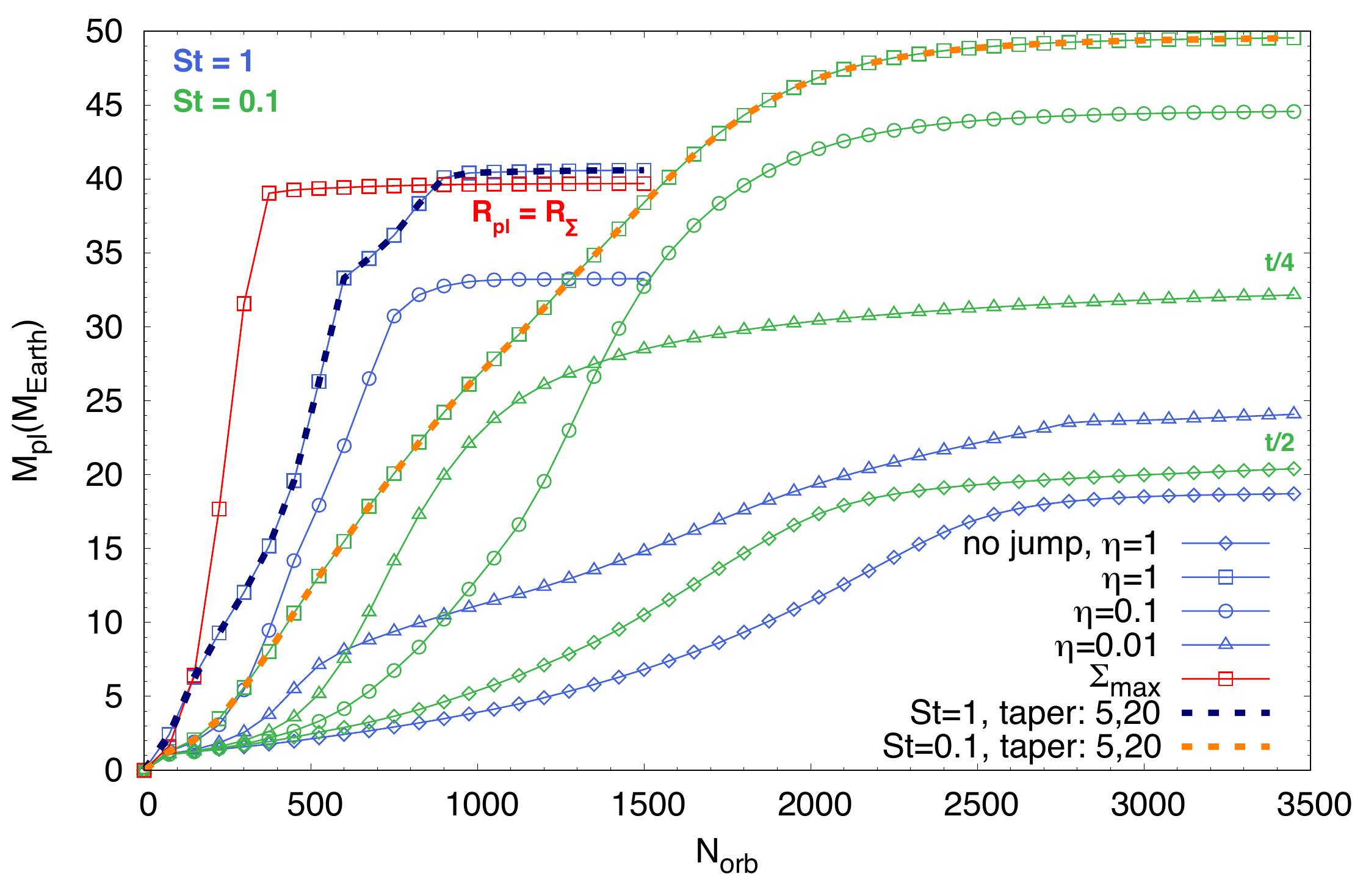}
    \caption{Evolution of the core's mass as the function of time. Two different Stokes numbers are modelled, $\mathrm{St}=0.1$ and $1$, shown with green, and blue colours, respectively. Diamonds are for the models, where no pressure maximum is present and $\eta=1$. For a given Stokes number the accretion efficiency $\eta$ is denoted by different symbols: squares represent $\eta = 1$, circles $\eta = 0.1$, and triangles $\eta = 0.01$. The red curve corresponds to the case in which the core is placed to the gas surface density maximum. If a certain curve is marked by "t/2" or "t/4" the time corresponding to a mass value should be multiplied by 2 or 4, respectively. Mass evolution curves for different tapering times are displayed by dashed curves. For a given Stokes number the two curves showing the core’s mass evolution with tapering time 5 and 20 periods are displayed with the same colour because these curves practically overlap each other.}
    \label{fig:fig_1}    
\end{figure}

\begin{figure*}
    \includegraphics[width=\textwidth]{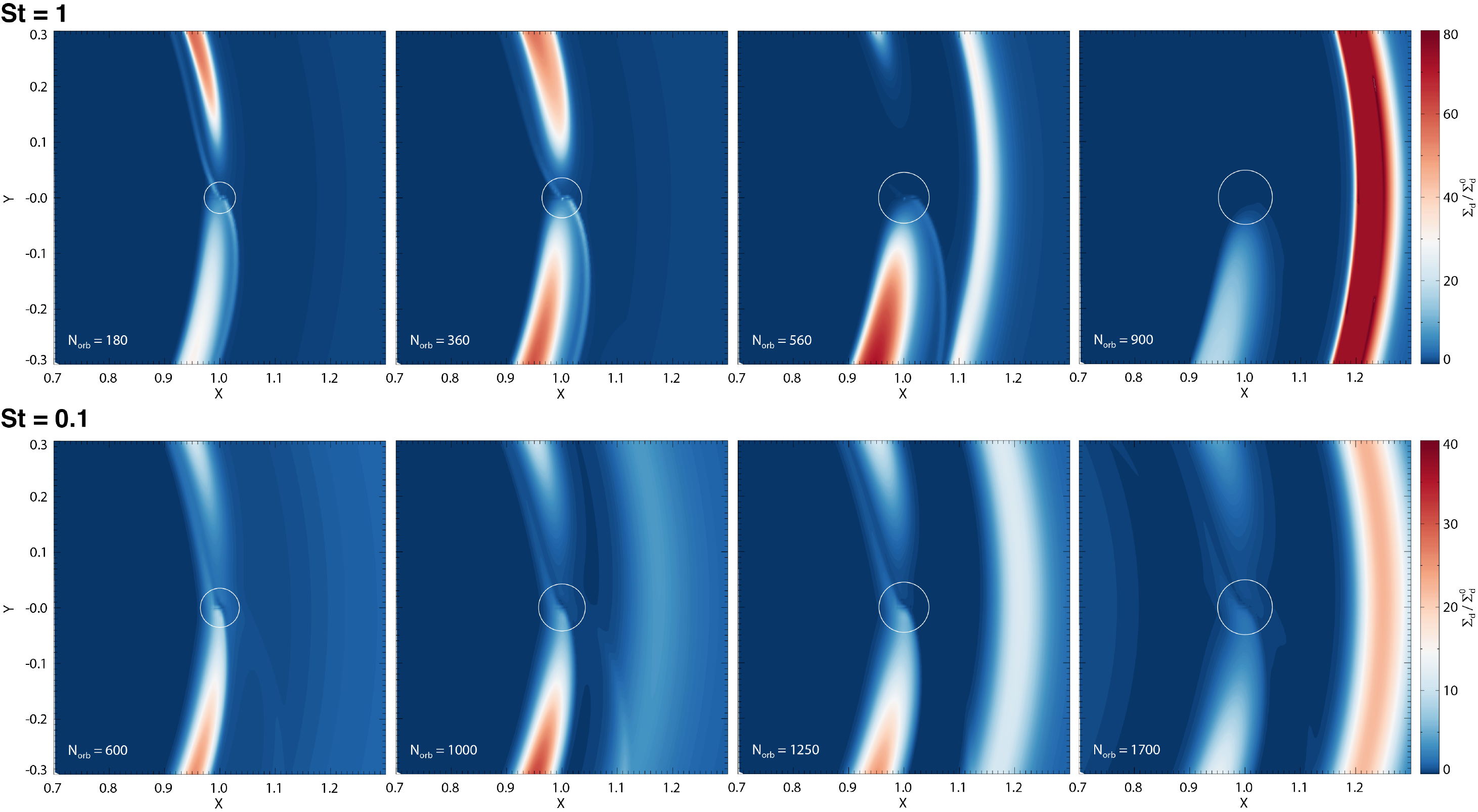}
    \caption{Normalised pebble surface density snapshots belonging to different phases of pebble accretion around the growing core. In the upper four panels the accretion of pebbles with Stokes number $\mathrm{St}=1$, while in the lower panels the accretion of pebbles with Stokes number $\mathrm{St}=0.1$ are displayed. White circles correspond to the core's Hill sphere. Movies for the complete evolution of pebble surface density in the vicinity of the core are available online.}
    \label{fig:fig_2}    
\end{figure*}
\begin{figure}
    \includegraphics[width=\columnwidth]{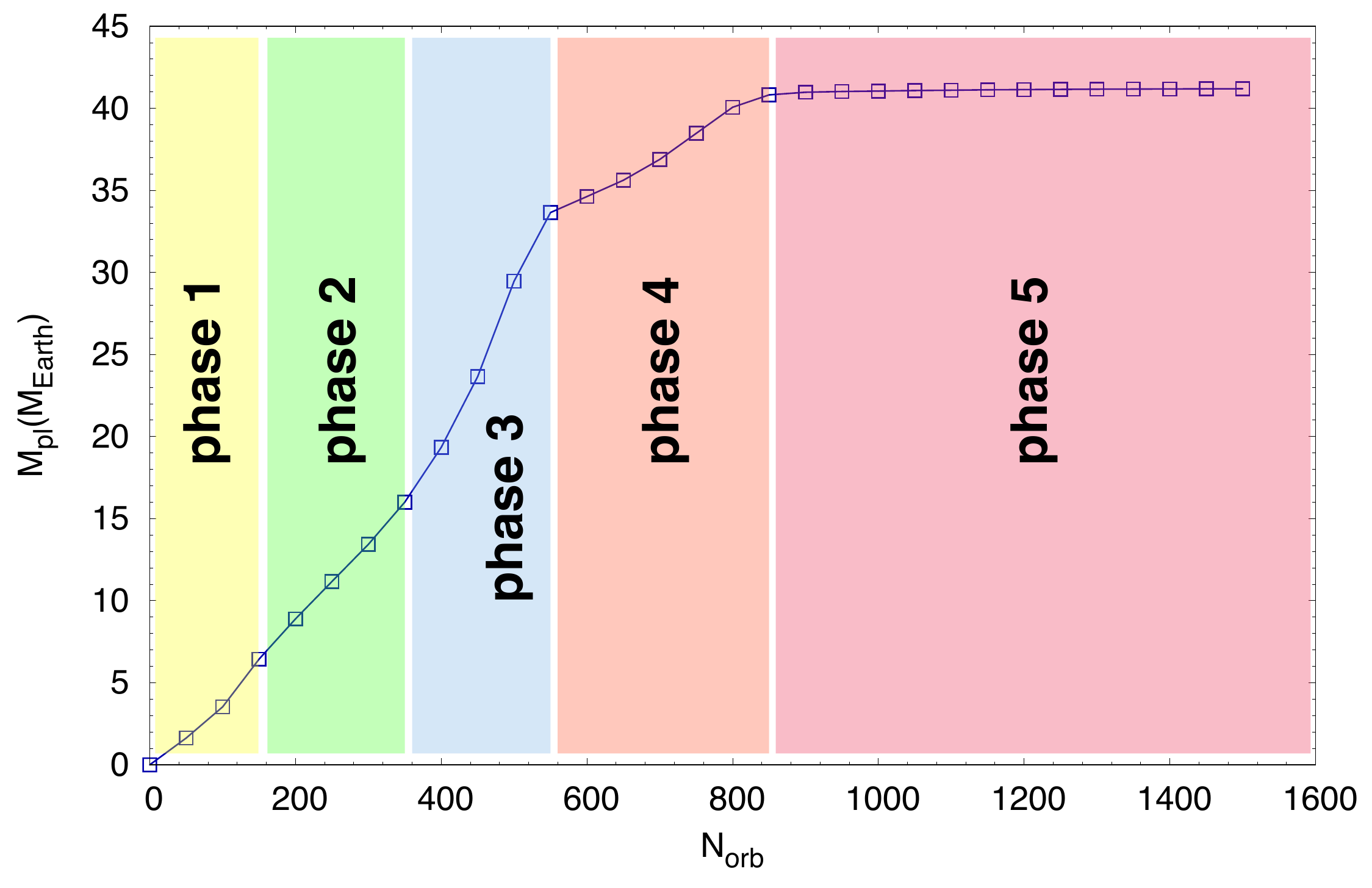}
    \caption{Growth of the planetary core as the function of time when $\mathrm{St}=1$ and $\eta=1$. The phases of growth are shaded with different colours.}
    \label{fig:fig_3}    
\end{figure}

\section{Growth of the core and isolation masses} 
In order to reveal how the final mass of the core depends on the particle size and on the accretion efficiency, we performed simulations in several models. In Figure \ref{fig:fig_1} nine of them are displayed, seven with a pressure maximum in which the surface density is given by Equation \ref{eq:initgas} with $\Sigma_\mathrm{g}^\mathrm{bg}=10^{-3}$ and two without pressure maximum  using a constant gas surface density: $\Sigma_\mathrm{g}=\Sigma_\mathrm{g}^\mathrm{bg}=10^{-3}$.

The curves (except the red one) are coloured corresponding to the Stokes-number of particles. Blue curves are for pebbles having $\mathrm{St}=1$, green curves for pebbles with $\mathrm{St}=0.1$. For a given value of the Stokes number, the accretion efficiency $\eta$ is marked by different symbols: squares represent $\eta = 1$, circles $\eta = 0.1$, and triangles $\eta = 0.01$. Additionally, diamonds are for simulations without pressure maximum with $\eta=1$ accretion efficiency. The red curve is for pebbles with $\mathrm{St}=1$, and $\eta=1$, but instead of the pressure maximum $R_\mathrm{pmax}$, the core is placed to the gas surface density maximum $R_\Sigma$.

The most visible outcome of our simulations is that in a pressure maximum a core can grow larger than in models without such a maximum. In our specific models without pressure maximum we obtain isolation masses $M_\mathrm{iso}\approx 19 M_\oplus$ for pebbles with $\mathrm{St}=1$ and $M_\mathrm{iso}\approx 20 M_\oplus$ for pebbles with $\mathrm{St}=0.1$. This result fits well with that obtained by \citet{Lambrechts+2014} using 3D hydrodynamic simulations covering a computational domain with a similar extent to our setup. 

We investigate first the core growth in models with a pressure maximum in which the core accretes pebbles with Stokes number $\mathrm{St}=1$ (blue curves of Figure \ref{fig:fig_1}). One can see that the most efficient accretion happens (not surprisingly) for $\eta = 1$. In this case, the maximum core mass is about $M_\mathrm{iso}\approx 41 M_\oplus$. The core mass growth as the function of time is steep, and the saturation of the final core mass is reached after $\sim 900$ orbital period. A quite similar outcome we have if $\eta=0.1$. In this case, the saturation mass is smaller, being around $M_\mathrm{iso}\approx 32 M_\oplus$, but it is obtained in $\sim 1000$ orbital periods. By using an even smaller accretion efficiency, $\eta = 0.01$, the mass growth function is not so steep, and the saturation mass reached around $\sim 4000$ orbital period is $M_\mathrm{iso}\approx 25 M_\oplus$ (not visible in Figure \ref{fig:fig_1}). This behaviour can be explained by the fact that the less efficient accretion allows pebbles to be drifted radially across the region of co-rotation due to drag without being accreted by the growing core. The saturation of the final mass is also present in this case, so when the core is reaching the critical mass, no pebbles could increase its mass. 

When the core grows by accreting pebbles with Stokes numbers $\mathrm{St}=0.1$ the curves of the mass growth are smoother and less steep when comparing them to the previous cases with pebbles having Stokes numbers $\mathrm{St}=1$. Interestingly, the pebble isolation masses are also significantly higher than in the case of $\mathrm{St}=1$ particles. The smoother curves of the core's mass growth are due to fact that the drag induced radial drift is slower for smaller pebbles, moreover, these particles are also more sensitive to turbulent diffusion. The less steep mass growing curves indicate a lower pebble accretion rate for the growing core because due to the slower radial drift, the relative velocities between the core and pebbles are smaller, too. One reason for the larger pebble isolation masses is that particles with smaller Stokes numbers are more sensitive to turbulent diffusion therefore they can pass easier across the pressure maximum at the outer gap edge. Moreover, during the longer formation time, more pebbles are accumulated in the core's co-rotation region because small pebbles are less sensitive to the drag induced radial drift. The maximum isolation mass that a core can reach is about $M_\mathrm{iso}\approx 50 M_\oplus$, when $\eta=1$, and $M_\mathrm{iso}\approx 45 M_\oplus$ when $\eta=0.1$ both with $\sim 3000$ orbital period saturation times. When $\eta=0.01$, the core's mass is growing very slowly reaching $M_\mathrm{iso}\approx 33 M_\oplus$ at the end of simulations at $T=14000$ orbital period but without reaching saturation. In this case, the formation of the pressure maximum at the gap's outer edge needs a much longer time than in cases of larger accretion efficiencies when the core grows faster. Once the pressure maximum at the gap's outer edge is formed, even if a certain amount of pebbles could penetrate through it by turbulent diffusion, due to the reduced accretion rate they are rather drifted inward than being accreted by the core. This result indicates that the pebble isolation mass depends on the balance between the core's accretion rate and the efficiency of radial drift in emptying the region of co-rotation. The above argumentation is also valid when explaining the fact that reduced accretion efficiencies in general result in lower pebble isolation masses. The loss of pebbles (with $\mathrm{St}=1$) when the accretion efficiency of the core is reduced ($\eta=0.1$) can be followed on a movie that is available in the online supplementary material: St-1-eta-01-Pmax.mpg.

As mentioned before, the dependence of the pebble isolation mass on the initial position of the growing core is also investigated for pebbles with $\mathrm{St}=1$. We have found that the maximum pebble isolation mass can be achieved when the core is placed exactly to the pressure maximum's position $R_\mathrm{c} = R_\mathrm{pmax}$. If the core is placed either to the surface density maximum, that is $R_\mathrm{c}=R_\mathrm{pmax}+\delta$, or to a position where $R_\mathrm{c} = R_\mathrm{pmax} - \delta$, where $\delta=R_\Sigma-R_\mathrm{pmax}$ the core's pebble isolation mass will be smaller than in the $R_\mathrm{c} = R_\mathrm{pmax}$ case, because in both cases the amount of pebbles in the vicinity of the core is less than when the core is at $R_\mathrm{pmax}$, moreover a significant amount of pebbles is also lost from the region of core's co-rotation. When investigating the red curve of Figure \ref{fig:fig_1}, one can see that initially the core's pebble accretion rate is higher (the mass growth curve is steeper) that is due to the larger relative velocities between the pebbles and the core. On the other hand, the core's final mass is smaller in this case, since a certain amount of pebbles are not incorporated in the region of the core's co-rotation any longer, as being drifted inward quickly when the pressure maximum vanishes. The movie on the pebble accretion (with $\mathrm{St}=1$ and $\eta=1$ parameters), when the core is placed at the gas surface density, is available in the online supplementary material: St-1-eta-1-densmax.mpg.

The process of mass growth has been analysed in detail for two cases, when $\mathrm{St=1}$ and $\mathrm{St=0.1}$ with accretion efficiency $\eta = 1$. The two movies showing the 2D pebble surface density evolution at the vicinity of the growing core are available in the online supplementary material: St-1-eta-1-Pmax.mpg, St-01-eta-1-Pmax.mpg.

First, we study the core's mass growth in the particular case when $\mathrm{St}=1$ and $\eta=1$. In the mass growth curve, a few well-defined breaking points can be identified, when the pebble accretion rate changes abruptly. These breaking points cannot be found when the core grows by accreting smaller pebbles with $\mathrm{St}=0.1$. Thus, the growing phases being discussed in what follows may not be regarded as general ones and may not be found in the other simulations.

Four snapshots of the pebble surface density in the vicinity of the core are shown in the upper panel of Figure \ref{fig:fig_2}, while the mass of the core (as the function of time) is displayed in Figure \ref{fig:fig_3}. We recall that in this particular case, the final mass of the core is $M_\mathrm{iso}\approx 41 M_\oplus$. There are four points in the mass growth curve, where the growth rate of the core changes, namely at $T\approx 160$, $T\approx 360$, $T\approx 560$, and $T\approx 900$ orbital periods. When $T\leq 160$, the core is fed directly by pebbles already accumulated in the pressure maximum forming a ring-like structure. When $T\approx 160$, this ring becomes discontinuous at the core's position, taking the well-known horseshoe shape structure (whose lower and upper branches are visible in Figure \ref{fig:fig_2}) while the feeding of the core by pebbles happens only from the lower branch. This kind of accretion occurs in the interval $160\le T \le 360$, and since the upper branch is not connected to the Hill-sphere of the growing core, the surface density of pebbles in the upper branch increases rapidly. Meantime, the Hill-sphere (the white circle) of the core grows with its mass, and eventually reaches the upper branch, while being steadily connected with the lower branch, too, see the second panel of Figure \ref{fig:fig_2}. Thus in the interval $360\le T \le 480$, the core's growth happens from both branches, resulting in an increased pebble accretion. (We note that at $T\approx 480$ a slight change in accretion happens.) Already around $T\approx 500$ the core reaches a critical mass $M_\mathrm{crit} \approx 25 M_\oplus$ that results in forming a pressure maximum at the gap's outer edge. Due to the turbulent diffusion, at this stage, this pressure maximum is not an insurmountable barrier for pebbles. However, as the core grows, the pressure maximum at the outer edge of the gap traps pebbles more efficiently, and at $T \approx 560$ cuts definitively the influx of pebbles from the outer disc, see the ring of pebbles outside of the core's orbit on the third panel of Figure \ref{fig:fig_2}. Thus in the interval $560 \le T \le 900$, the pebble accretion takes place from the pebbles accumulated originally in the pressure maximum in which the core is formed, which region becomes gradually the co-orbital region of the growing core. Having accreted the majority of the solid material by the core from this region at $T\approx 900$, the core reaches its isolation mass, see the fourth panel of Figure \ref{fig:fig_2}, though some amount of pebbles also leaves the co-orbital region, presumably by an inward drift. It is noteworthy that during the core growing process, the density maximum in the gas gradually vanishes due to the gap opening. 

In the lower panel of Figure \ref{fig:fig_2} four snapshots of the pebble surface density are displayed when $\mathrm{St}=0.1$ and $\eta=1$. The phases separated by the breaking points described previously for pebbles with $\mathrm{St}=1$ cannot be identified in this case. However, the main characteristics of pebble accretion can still be identified as (i) the formation of the discontinuity in the ring-like structure at the core ($T\approx 600$), (ii) formation of the pressure maximum at the outer edge of the gap opened in the gas by the core ($T\approx 960$), (iii) accretion of pebbles mainly from the co-orbital region of the core but also pebbles passing through the pressure maximum by diffusion ($T\approx 1100$), and finally, (iv) the emptying of the solid material from the co-orbital region $T\approx 2500$. Having cut the influx of pebbles by the pressure maximum at the gap's outer edge, a significant amount of solid material is already accumulated in the core's co-orbital region that is mainly accreted by the core since smaller pebbles are not subject to very fast radial drift.

In both cases, the ring-like structure appears, when the core's mass is $M_\mathrm{core}\approx 25 M_\oplus$. However, due to the turbulent diffusion, pebbles with smaller Stokes number $\mathrm{St}=0.1$ can also pass across the pressure maximum developed at the gap's outer edge more easily than pebbles with Stokes number $\mathrm{St}=1$, modestly contributing to the growth of the core at this stage. Due to the longer formation time, a larger mass of pebbles with Stokes number $\mathrm{St}=0.1$ can accumulate in the core's co-rotation region than in the $\mathrm{St}=1$ case. Thus since smaller pebbles are less sensitive to the radial drift than the larger ones, the final mass of the core, $M_\mathrm{iso}\approx 50 M_\oplus$, is larger in the $\mathrm{St}=0.1$ case, too.

We also run simulations with tapering times being 5 and 20 orbital periods of the core both for pebbles with $\mathrm{St}=0.1$ and $\mathrm{St}=1$ Stokes numbers to see the possible dependence of the pebble isolation mass on the initial dust reservoir that accumulates during the tapering time. Surprisingly, we have found in all of these simulations that the tapering time does not affect the pebble isolation mass, see the dashed curves in Figure \ref{fig:fig_1}, where the orange curve displays the $\mathrm{St}=0.1$ cases, and the dark blue curve the $\mathrm{St}=1$ cases. We note that for a given Stokes number the curves showing the core's mass evolution are displayed with the same colour because these curves practically overlap each other. The above results indicate that even with the maximum accretion efficiency the growing core can only accrete a limited amount of pebbles accumulated in the pressure maximum. Pebbles that are not accreted by the growing core until the time when the original pressure maximum vanishes, leave the region of core's co-rotation by radial drift. The final pebble isolation mass is then the result of a balance between the pebble accretion efficiency of the growing core, the radial drift speed of pebbles through the co-rotation region, and the penetrability of the pressure maximum at the gap's outer edge by the pebbles.

\section{Discussion and conclusion}

The growth of a solid core up to tens of Earth masses is an essential ingredient of giant planet formation in the core accretion paradigm. When the core's mass approaches the pebble isolation mass, pebble accretion ceases. If the core's mass is large enough, the gaseous envelope of the core collapses, and a giant planet forms. 

For this reason, pebble isolation mass has been thoroughly studied in protoplanetary discs characterised with a power-law surface density profile. When giant planet formation takes place in a pressure maximum of a protoplanetary disc, pebble accretion has been investigated only in a very few works. In a recent study of \cite{Guilera&Sandor2017}, the role of pebble accretion has been investigated in the formation of giant planets in the pressure maxima of protoplanetary discs resulting in a very efficient and rapid formation. More recently, pebble accretion in a generic pressure maximum has been studied by \citet{Morbidelli2020} showing that in the absence of outer flux of pebbles, a giant planet can be formed with a smaller mass core. The effect of the pebble isolation mass has not been taken into account in any of the above works. Based on the preliminary results of this work, \cite{Guilera+2020} have also applied a larger limit for pebble isolation mass than the typical value for power-law discs. It has been found that a larger isolation mass results in the formation of a larger core, moreover, the giant planet is formed in a short time. 

Motivated by the efforts that aim at studying planet formation in pressure maxima of protoplanetary discs, we investigate how the pebble isolation mass changes in a generic pressure maximum when comparing its value to that obtained in a disc without pressure maximum. We consider a planetary core that forms in a pressure maximum from pebbles accumulated there and grows further by accreting the already accumulated pebbles, and also pebbles being drifted from the outer disc until reaching the core's pebble isolation mass. We note that the isolation masses found in this research certainly depend on the physical properties of the pressure maximum described by Equation \ref{eq:initgas}. In future work we are planning to study this dependence in a more detailed way.

We also run simulations with a constant surface density profile, in which we find $M_\mathrm{iso}\approx 19 M_\oplus$ being in excellent agreement with the results of \cite{Lambrechts+2014}. In the generic pressure maximum considered, we have found much larger isolation masses, namely, $M_\mathrm{iso}\approx 41 M_\oplus$ for $\mathrm{St}=1$, and $M_\mathrm{iso}\approx 50 M_\oplus$ for $\mathrm{St}=0.1$ particles. Investigating the pebble accretion process of the growing cores (see the attached movies), we identify a few reasons being responsible for the increased isolation masses. In the generic pressure maximum, that is associated with a density maximum in gas, the gap opening by the core requires larger core mass, $M_\mathrm{crit}\approx 25 M_\oplus$, than in the case of a disc with a power-law surface density profile. Due to the turbulent diffusion of the gas, the pressure maximum emerging at the gap's outer edge is initially penetrable with a certain rate for pebbles. The penetrability of this pressure maximum depends on the pebble sizes, because due to the turbulent diffusion pebbles with Stokes number $\mathrm{St}=0.1$ can pass across the pressure maximum easier than pebbles with $\mathrm{St}=1$. When the influx of pebbles stops, the core's mass can still grow by accreting pebbles incorporated in the core's co-rotation region. However, this pebble population may not be entirely accreted by the core, because some amount of it leaves the region of co-rotation by radial drift. Thus the final pebble isolation mass of a planetary core formed in a pressure maximum is set by different physical phenomena, such as the diffusion of pebbles across the pressure maximum developed at the outer edge of the gap, the accretion efficiency of the core and the speed of the drag-induced radial drift of pebbles that empty the core's co-rotation region. Regarding the core's pebble isolation mass, it is of high importance the mass of pebbles accumulated first in the pressure maximum, and later on in the region of co-rotation. A core that is accreting pebbles with Stokes number $\mathrm{St}=0.1$ grows more slowly than by accreting pebbles with $\mathrm{St}=1$. Thus the time needed for gap opening by the core is also longer in the $\mathrm{St}=0.1$ case. During the longer formation time, a larger mass of pebbles is accumulated initially in the pressure maximum and later on in the core's co-rotation region. The radial drift for pebbles with $\mathrm{St}=0.1$ is slower, thus the core's final mass is also larger in this case. Summarising the above findings, a core that is accreting pebbles with $\mathrm{St}=0.1$ grows larger, while a reduced accretion efficiency results in a smaller pebble isolation mass.

Another effect that could modify the isolation mass is the width of the feeding zone of the pressure maximum, which is the region from where pebbles can arrive at the core's location. If there are multiple pressure maxima that eventually lye close to each other, the mass of the core may not reach the isolation mass found in our research. 

We can finally conclude that pebble isolation mass can be high in a generic pressure maximum of a disc resulting in a massive solid core and short formation time of a giant planet if the influx of solids is guaranteed during the process of giant planet formation. 

\section*{Acknowledgements}
This research is supported by the Hungarian National Research, Development, and Innovation Office (NKFIH), under the grant K-119993. The authors thank the constructive discussions and suggestions to O. Guilera and an anonymous referee, whose suggestions helped us to improve our manuscript considerably.

\section*{Data availability}
The data underlying this article are available in the article and in its online supplementary material.




\bibliography{sandor_regaly}








\bsp	
\label{lastpage}
\end{document}